# Knock Intensity Distribution and a Stochastic Control Framework for Knock Control

**Mateos Kassa, Carrie Hall, and Michael Pamminger**

*Illinois Institute of Technology*

**Thomas Wallner**

*Argonne National Laboratory*

**Abstract**

One of the main factors limiting the efficiency of spark-ignited engines is the occurrence of engine knock. In high temperature and high pressure in-cylinder conditions, the fuel-air mixture auto-ignites creating pressure shock waves in the cylinder. Knock can significantly damage the engine and hinder its performance; as such, conservative knock control strategies are generally implemented that avoid such operating conditions at the cost of lower thermal efficiencies. Significant improvements in the performance of conventional knock controllers are possible if the properties of the knock process are better characterized and exploited in knock controller designs. One of the methods undertaken to better characterize knocking instances is to employ a probabilistic approach, in which the likelihood of knock is derived from the statistical distribution of knock intensity.

In this paper, it is shown that knock intensity values at a fixed operating point for single fuel and dual fuel engines are accurately described using a mixed lognormal distribution. The fitting accuracy is compared against those for a randomly generated mixed-lognormally distributed data set, and shown to exceed a 95% accuracy threshold for almost all of the operating points tested. Additionally, this paper discusses a stochastic knock control approach that leverages the mixed lognormal distribution to adjust spark timing based on knock intensity measurements. This more informed knock control strategy would allow for improvements in engine performance and fuel efficiency by minimizing knock occurrences.

**Introduction**

Engine knock is the inadvertent auto-ignition of fuel in localized high pressure and temperature regions inside the cylinder [1,2]. Engine knock can result in significant engine damage and marks one of the main obstacles preventing spark-ignited (SI) engines from attaining higher fuel efficiencies. Various in-cylinder conditions such as high residual concentration or high temperature and pressure are prone to knocking and an adjustment to the fueling or ignition strategy is essential to prevent or minimize knock events at such conditions.

The conventional approach to avoiding knock in SI engines consists of delaying the combustion phasing by retarding the spark timing [3]. A combustion event occurring later in the combustion stroke (further away from top dead center) has a lower tendency to knock since the combustion pressure and temperature are lower. However, delaying the combustion phasing also reduces the fuel efficiency since less work can be extracted by the late combustion [4].

The tendency of the fuel to auto-ignite is not only dependent on the in-cylinder conditions, but also on the knock resistance (RON-research octane number) of the fuel. As such, increasing the fuel's octane rating can help engines avoid knock without compromising fuel efficiency. Since higher octane fuels are more expensive, there has been interest in using high octane fuels in a dual-fuel combustion strategy. Engines with dual-fuel capabilities can use a low RON fuel and a high RON fuel simultaneously to optimize the fuel mixture's knock resistance by controlling the proportion of each injected fuel. A number of studies have explored the implementation of a dual-fuel strategy to suppress knock [5-8].

In both conventional and dual-fuel SI engines, controlling the spark timing and/or the fuel split is challenging since several engine operating parameters can play a role in knock occurrences. Accurately estimating the knock propensity of a combustion cycle and determining the appropriate spark timing or knock resistance of the fuel is an essential aspect of a SI engine. A closed-loop knock control system is generally implemented in production vehicles, in which the spark timing is continuously adjusted based on a measurement of vibrations recorded by an accelerometer [3]. Similarly, the dual-fuel SI engine leverages measurements of the combustion intensity to adjust the fuel split control. Engine manufacturers generally employ highly conservative knock control approaches to avoid knocking instances (at the cost of fuel efficiency losses). These approaches are favored because knocking instances are seemingly randomly occurring and accurately characterizing the knocking propensity of a combustion cycle proves difficult.

Prior work has shown that knock intensity and knock events have characteristic statistical properties that can be leveraged for control purposes [9-12]. Spelina et al. [10,11] explored and quantified the statistical characteristics of the knock process across a broad range of engine operating conditions. Their studies showed that, for the operating points they studied, knock behaved to a good first approximation as a cyclically independent random process. Furthermore, in [11], knock intensity (KI) was fitted as a gamma and lognormal distribution. The study concluded that knock data did not in fact conform to either gamma or lognormal distribution; however, the study suggested that a lognormal model would be an adequate approximation for control purposes.

Studies by Peyton Jones et al. [13-15] concluded that significant improvements in the performance of standard knock controllers may be possible if the properties of the knock process are better characterized and exploited in knock controller designs. Several studies have indeed shown that a more stochastic approach of knock control considering the statistical properties of combustion intensity can offer better spark timing control (and thereby better knock control) [16-18]. To the authors' best knowledge, no such studies have been conducted on the development of a stochastic fuel-split knock control strategy for dual-fuel SI engines.

In order to develop improved stochastic knock control strategies for these dual-fuel engines, it is essential to have an accurate representation of the statistical properties of knock and knock intensity. Dual-fuel engines enable the study of knock from an advanced engine framework; hence, characterizing knock from these events has value for future work on high efficiency engines. This paper discusses the investigation of statistical properties of knock that can be leveraged for an effective knock control strategy to reduce efficiency losses (and/or reliance on high RON fuel) while simultaneously avoiding knock. Lognormal and mixed lognormal distributions are investigated to represent knock intensity in both single fuel and dual fuel cases. Unlike the lognormal distribution, a mixed-lognormal distribution is shown to accurately (with 95% confidence) characterize knock intensity across a wide range of operating conditions. A more accurate characterization of knock could be leveraged to more accurately simulate knock occurrence in high-fidelity engine models and also could be leveraged in stochastic control methods.

In the next section, the engine used for this study is described along with details of the range of data utilized for the analysis of knock intensity. Subsequently, the approach for calculating knock intensity is presented followed by discussion of the characterization of the distribution of KI and approaches undertaken to accurately model the distribution of knock. The last section of this paper discusses one possible application of these improved knock intensity characterizations in a knock control strategy that leverages the mixed-lognormal distribution.

**Experimental Setup and Data Collection**

The statistical analysis of KI was conducted using data from a series of steady state tests on a single-cylinder engine that was equipped with both a port-injection system and a direct-injection system to allow for both single fuel and dual fuel injection strategies. The engine configuration resembles that of a modern gasoline direct-injection (GDI) engine and the specifications of this engine are detailed in Table 1. The same engine was used in [19-21].

Table 1. Single Cylinder Engine Specification

| Parameter | Value |
| --- | --- |
| Displacement Volume | 0.6264 L |
| Number of Cylinders | 1 |
| Stroke | 100.6 mm |
| Bore | 89.04 mm |
| Compression Ratio | 10.5:1 and 12.5:1 |
| Number of valves | 4 |
| Spark Plug | NGK, 0.7 mm gap |

Fresh air was supplied to the engine from an Atlas Copco air compressor and was maintained at appropriate levels during throttled conditions using a Parker pilot operated regulator in the intake. Cylinder pressure on this engine was recorded using an AVL GU21C transducer and AVL 365X crank angle encoder.

The engine used for this study was equipped with a liquid port-injection system manufactured by Ford, which operates at a 4.1 bar gauge injection pressure. Gaseous port-injection is done using a Bosch injector, which operates at 7 bar gauge injection pressure. In addition, this engine also has a natural gas direct-injection system. This gaseous direct-injection system uses a Delphi injector with an outward opening valve and a maximum injection pressure of 15 bar gauge.

The steady state tests consisted of operations at two compression ratios, namely 10.5:1 and 12.5:1. Additionally, variations of the engine load were introduced by running the engine at part load, wide-open throttle, and boosted conditions (IMEP varying from 5.5 bar to 20 bar). Furthermore, a variation of fuels and fuel combinations were used at each operating point. The variations of fuels considered in this study include compressed natural gas (CNG) and three gasoline type fuels, namely E10, E85, and EEE. Different blends and injection of the fuels were considered; additional details about the fuels and fueling strategy are provided in [19]. Lastly, a sweep of spark timing was used to vary the knocking state in the operating conditions. The engine speed throughout these tests is fixed at 1500 RPM. A summary of the operating points is illustrated in Table 2.

Table 2. Summary of Operating Parameters

| Operating Parameter | Min | Max |
|---|---|---|
| Compression Ratio [-] | 10.5 | 12.5 |
| IMEP [bar] | 5.6 | 20.0 |
| Spark Timing [°BTDC] | -6 | 58 |
| Fuel Blends | Blend Ratio [% mass] | |
| E10 | - | |
| CNG | - | |
| E85 | - | |
| E10-CNG | 25/50/75 | |
| E10-EEE | 25/50/75 | |

In total, a set of 247 different operating points was considered. For each operating point, a set of 3 tests were recorded each consisting of 372 cycles (for a total of 1116 cycles per operating point). The dataset represents a wide range of knock intensities and knock behaviors and was determined to be suitable for meaningful statistical analysis. The next section discusses the process undertaken to calculate KI from in-cylinder pressure measurements as a prelude to the statistical analysis that follows.

**Knock Intensity Calculation**

KI can be calculated from in-cylinder pressure traces or accelerometer signals. Both methods have been shown to yield similar results and essentially capture the ringing intensity of the combustion cycle [9]. In this study, the statistical analysis of KI was derived from the pressure trace. However, similar results are also expected from accelerometer based KI measurements.

A three step process was used to calculate KI from a pressure trace. First, a crank angle window is selected where a knock event is anticipated. This window is generally determined with respect to the spark timing. In this study, the window for the analysis is defined as spark timing+20°CA (crank angle) to spark timing+110°CA. Second, the pressure trace in that window is filtered through a range of frequencies corresponding to knock (based on the speed of sound in the cylinder and the resonances frequencies consistent with the cylinder geometry). For this engine, the pressure trace is filtered in the frequency range of 3 kHz to 25 kHz, consistent with the dimensions of the engine and the literature [20, 22, 23]. Finally, the filtered pressure trace is used to generate a single metric defined as KI – this value is generally either the area underneath the redressed signal or the maximum value of the redressed signal. For this study, the maximum value of the redressed signal (absolute value) in the knock window is chosen as the KI metric. Additional information on the use of KI can be found in [24].

Figure 1 illustrates both the pressure trace and the filtered pressure of a typical combustion cycle. As shown, high amplitude oscillations are observed in the region corresponding to the combustion period. The highest amplitude (defined here as KI) is selected for each of the 1116 cycles at the operating point and analyzed to determine the knocking conditions of that operating point.

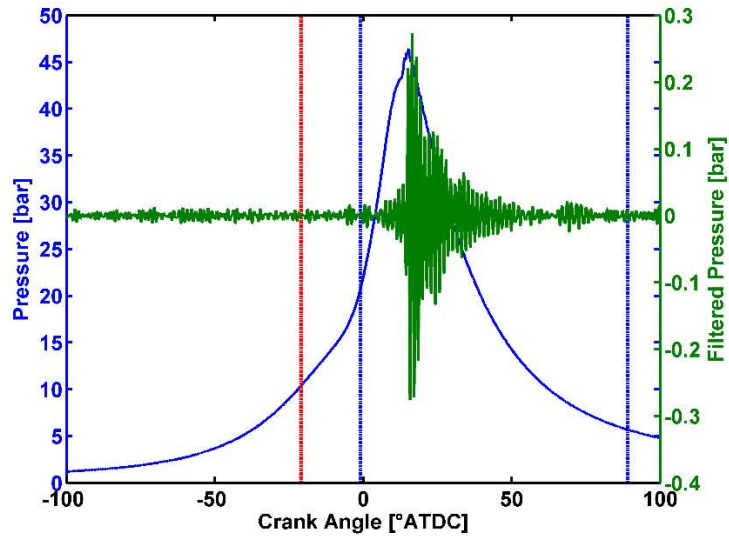

Figure 1. Illustration of the pressure trace and filtered pressure trace (3-25kHz) for a sample combustion cycle. The red line indicates spark timing; the blue lines represent the knock window.

Figure 2 illustrates the evolution of the filtered pressure trace amplitude for different operating points (with varying knock states). In each of the subfigures, all 1116 filtered pressure traces corresponding to each cycle of the operating point are plotted simultaneously. The comparison of the filtered pressure trace at different spark timings illustrates the tradeoff between optimal combustion phasing and knock. As spark timing is incrementally advanced from 15 °BTDC to 21 °BTDC, more favorable combustion phasing is achieved, yet the in-cylinder vibrations due to the severity of the combustion are more significant and potentially harmful to the engine. Higher amplitude of the filtered pressure trace and higher KI values are recorded at the earlier spark timings.

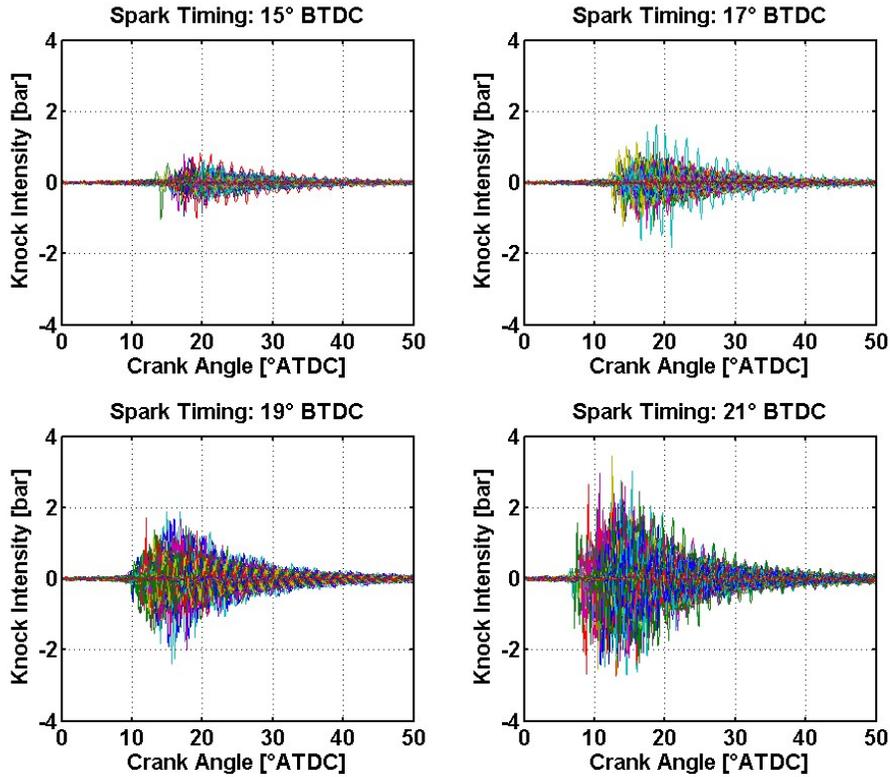

Figure 2. Filtered pressure trace amplitude evolution with spark timing

(E10 - 1500RPM / 7bar IMEP)

The impact of spark timing on KI values over the 1116 cycles can be more clearly seen in Figure 3, which illustrates the evolution of the average KI with spark timing. As spark timing occurs earlier, the cycle-to-cycle variations in KI increase making knock control difficult (even in steady-state conditions). The high cyclic variations are indicated by the $5^{th}$ and $95^{th}$ percentile bars, which show a high range of KI values at fixed operating points.

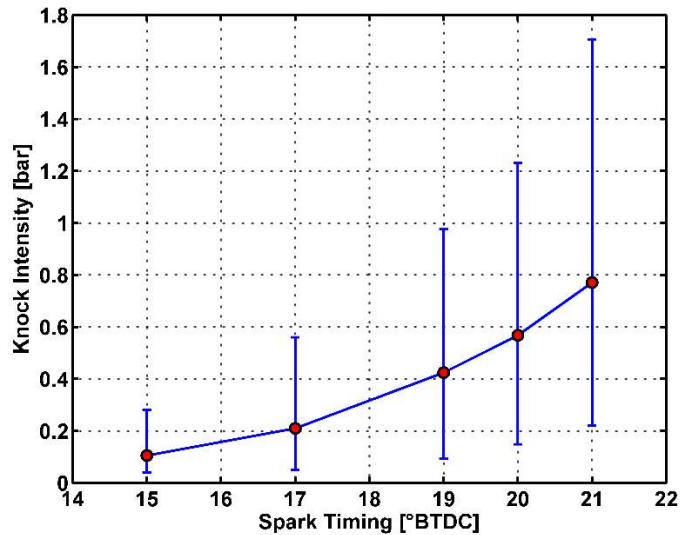

Figure 3. Evolution of average knock intensity with spark timing. The error bars represent the 5th and 95th percentile. (E10 - 1500RPM / 7bar IMEP)

As knock behavior is affected by many aspects of the combustion process, slight differences in the gas exchange and combustion process can have a significant impact on the resulting knock intensity. The concentration of residual gas, the temperature of the intake charge, the mixing level of the air-fuel mixture in the cylinder, the local temperatures and composition in the cylinder are among some of the important aspects of a combustion cycle that determine the level of knock in a combustion cycle. Although each mentioned parameter might only vary slightly from cycle-to-cycle at a fixed operating point, their combined effects make knock behavior starkly different between cycles.

Considering the large cyclic variations, the knocking condition for a specific operating point is better represented by a probability distribution of KI (instead of the average value

Hall 11 DS-17-1596

for example). The complete characterization of KI by its probability distribution, however, presupposes that the KI is an independent variable with, in this case, no cyclic dependence. The next section demonstrates analytically the cyclic independence of KI. The autocorrelation function is used to indicate that little or no correlation exists between the KI values, which supports the consideration of probabilistic distributions to better capture knock intensity.

**Cyclic Independence of Knock Intensity**

As discussed in the previous section, several factors simultaneously affect the occurrence of knock and the associated magnitude of the KI. As such, it is expected that KI will vary from cycle-to-cycle. This section aims to furthermore indicate that the KI values at a fixed operating point can be represented as an independent random process. Spelina et al. conducted similar studies on a single fuel engine and have showed that knock behaved (to a good first approximation) as a cyclically independent random process for the operating points they tested.

Here, despite the secondary fueling system and the varying fuel combinations employed, knock is still observed to exhibit very little cyclic correlation. The lag k autocorrelation function is calculated using Equation (1) below,

$$r_k = \frac{\sum_{i=i}^{N-k}(x_i - \bar{x})(x_{i+k} - \bar{x})}{\sum_{i=1}^{N}(x_i - \bar{x})^2} \quad (1)$$

where $N$ is the total number of points or 1116 in this case, $x_i$ represents the $i^{th}$ knock intensity in the operating point and $\bar{x}$ represents the mean knock intensity. The calculation of the autocorrelation is used to analyze the presence of any cyclic correlation in KI.

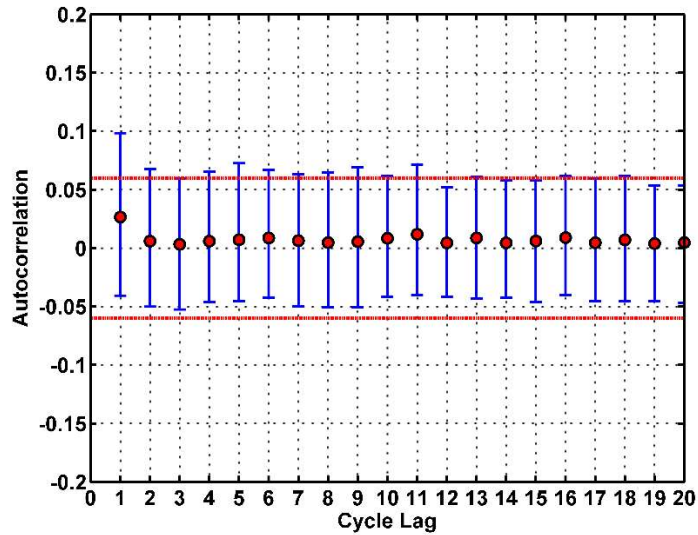

Figure 4. Normalized autocorrelation function of KI values

Figure 4 illustrates the normalized autocorrelation function (ACF) of all the sets of KI values obtained from the 247 operating points. The red dots in Figure 4 are used to mark the average autocorrelation values pertaining to each cycle lag. The error bars are used to identify the $5^{th}$ and $95^{th}$ percentile of the autocorrelation value at each cycle lag. The bounds shown by the two red lines indicate the 95% confidence region to reject the hypothesis of cyclic correlation. The upper and lower confidence bound are calculated based on the sample size.

As is evident from Figure 4, the ACF of KI for each cycle lag is very low and generally falls within the bounds confirming that KI is cyclically independent (with 95% confidence). Slightly higher ACF is observed for cycle lag 1, which suggests that subsequent cycles show more correlation in knock intensity. Furthermore, the range of ACF suggests that some operating points are likely to reflect higher levels of cyclic correlation than others. Nevertheless, the ACF is still significantly low enough to justifiably represent KI as a cyclically uncorrelated event.

**Modeling of Knock Intensity Distribution**

Considering the cyclic independence of KI at fixed operating points, KI can be fully characterized by the PDF or CDF of its distribution, as illustrated in Figure 5. Although a single KI value (ex. mean) or set of KI values might not be sufficient to fully characterize the knocking properties of the operating point, the PDF or CDF will provide complete information about the knock state and the associated KI values of that operating point. An understanding of the distribution of KI at different operating points and for different knock levels can be used as an effective tool to better interpret the measurements of KI. Knock control can be improved by leveraging a-priori knowledge of the distributions. One main goal of this paper is to identify a probability distribution that best characterizes the empirical PDF and CDF of KI at varying operating points.

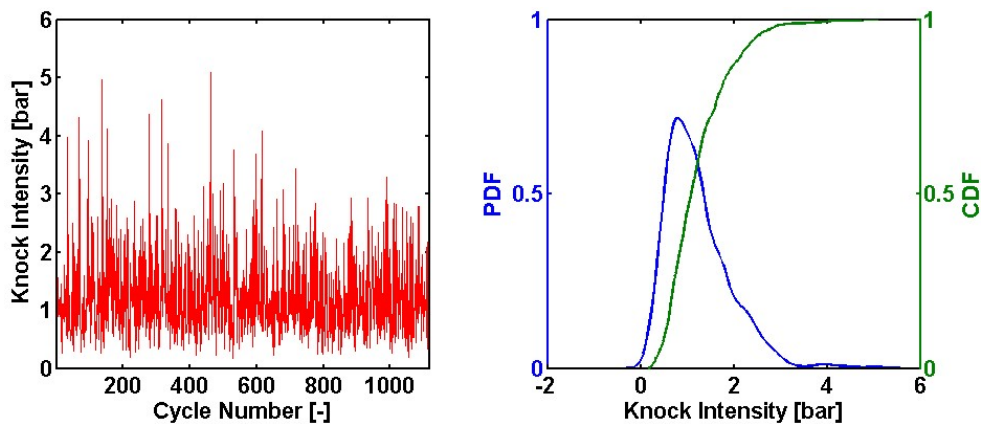

Figure 5. PDF and CDF representation of sample KI values

The lognormal distribution and mixed lognormal distribution are tested as potential representations of KI distribution for data from the dual fuel engine. The empirical CDF of KI is fitted using both these distributions. The accuracy of the fits is evaluated by comparing the empirical CDF and the model CDF. Two metrics are used to evaluate the accuracy of the fit: the coefficient of determination ($R^2$ value) and the Kolmogorov-Smirnov (KS) distance. The two metrics are calculated using equation (2) and (3) to compare the empirical CDF $Y$ with the estimated CDF $\hat{Y}$.

$$R^2 = \frac{\sum_{i=1}^{N}(Y_i - \bar{Y})^2 - \sum_{i=1}^{N}(Y_i - \hat{Y}_i)^2}{\sum_{i=1}^{N}(Y_i - \bar{Y})^2} \qquad (2)$$

$$KS = \sup|Y - \hat{Y}| \qquad (3)$$

A threshold of $R^2$ and KS distance values are determined to indicate the $R^2$ and KS distance values associated with a 95% confidence level. This is accomplished by fitting a large dataset of randomly generated data that are lognormally and mixed lognormally distributed. This approach of evaluating the fitting accuracy is the same method employed by Spelina et al. to reject the lognormal distribution as a potential indicator of knock events.

*Lognormal Distribution*

First, the lognormal distribution is used to fit the 247 sets of KI values (each consisting of 1116 data points). The PDF and CDF of a lognormal distribution are given by Equations (4) and (5) respectively,

$$f(x) = \frac{1}{x\sqrt{2\pi\sigma^2}} e^{-\frac{(\ln(x)-\mu)^2}{2\sigma^2}} \tag{4}$$

$$F(x) = \frac{1}{2} + \frac{1}{2}\text{erf}\left(\frac{(\ln(x)-\mu)^2}{2\sigma^2}\right) \tag{5}$$

where μ and σ are the mean and standard deviation of the variable's natural logarithm. The maximum likelihood estimates (MLE) of μ and σ² are provided by Equations (6) and (7) respectively.

$$\hat{\mu} = \frac{1}{N}\sum_{i=1}^{N} \ln x_i \tag{6}$$

$$\widehat{\sigma^2} = \frac{1}{N}\sum_{i=1}^{N} (\ln x_i - \hat{\mu})^2 \tag{7}$$

For each of the 247 operating points, $\hat{\mu}$ and $\widehat{\sigma^2}$ are determined using Equations (6) and (7), and the PDF and CDF are fitted for the range of KI values (using Equation (4) and (5)). The fitted lognormal distribution is compared against the empirical distribution of the data.

The fitting accuracy is illustrated in Figure 6, which shows three levels of fitting accuracy used to indicate the fitting accuracy of the worst cases, best cases, and median cases. In Figure 6, the distribution of the natural logarithm of KI is illustrated instead of that of KI to easily visualize the full distribution (minimizes skewing effect). The blue lines represent the probability density function (PDF) of the KI measurements, the green lines represent the empirical cumulative distribution function (CDF), and the dotted red lines represent the resulting estimation of the PDF and CDF from the lognormal fitting of KI (using $\hat{\mu}$ and $\widehat{\sigma^2}$ ) .

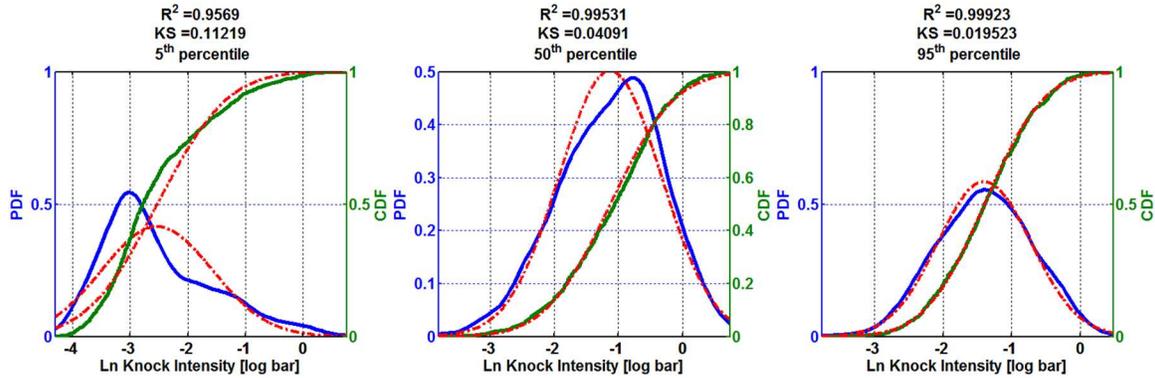

Figure 6. Lognormal fitting of KI with varying degrees of fitting accuracy

The leftmost figure in Figure 6 is an illustration of the lognormal fitting of KI from the operating point corresponding to the 5th percentile $R^2$ value among all fitting $R^2$ results (used to illustrate the 'worst case' fitting result). Similarly, the rightmost figure corresponds to the 95th percentile $R^2$ value (used to illustrate the 'best case' fitting result), and the center figure shows the fitting result of the operating point corresponding to the 50th percentile $R^2$ value (used to illustrate the median case fitting result). The fitting accuracy displayed no particular correlation with specific aspects of the operating conditions (load/spark timing/knock level). The fitting accuracy across the 247 operating points considered was seemingly random; Figure 6 is used to illustrate the range of fitting accuracy achieved with a lognormal fit of the KI distributions.

To further assess the fitting accuracy, the $R^2$ and KS values are compared against the expected $R^2$ and KS values from the fitting of truly lognormally distributed data. Using

MATLAB's random value generator, a set of 10,000 lognormally distributed data are generated (each comprising of 1116 points). Each data is fitted and the corresponding $R^2$ and KS distance values are calculated. The 5th percentile $R^2$ value was found to be 0.9985 and the 95th percentile KS distance was 0.283. Although a high value is desired for $R^2$, a high KS value is indicative of a large disparity between the empirical CDF and the fit. These values ($R^2$ = 0.9985 and KS = 0.283) represent the fitting accuracy expected (95% of the time) if the variable fitted is indeed lognormally distributed; as such, they are used to establish the 95% confidence level threshold ( solid red lines in Figures 7 and 8).

The fitting accuracy of KI from the 247 operating points is illustrated in Figure 7 and Figure 8, which display the $R^2$ and KS distance respectively. In both figures, the red line marks the threshold corresponding to 95% confidence level. The fitting accuracy of a lognormal fitting (for data points that are lognormally distributed) is expected to yield an $R^2$ value greater than 0.9985 and a KS distance lower than 0.0283 95% of the time if it is lognormally distributed.

However, the fitting accuracies are almost always below the expected level. Thus, it can be concluded that KI cannot be accurately described using a lognormally distributed data. However, the approximation of the distributions as lognormal can still be a useful control tool. This is also in agreements with the findings of Spelina et al. In the next section, it is shown that a mixed lognormal distribution captures the KI distribution more accurately.

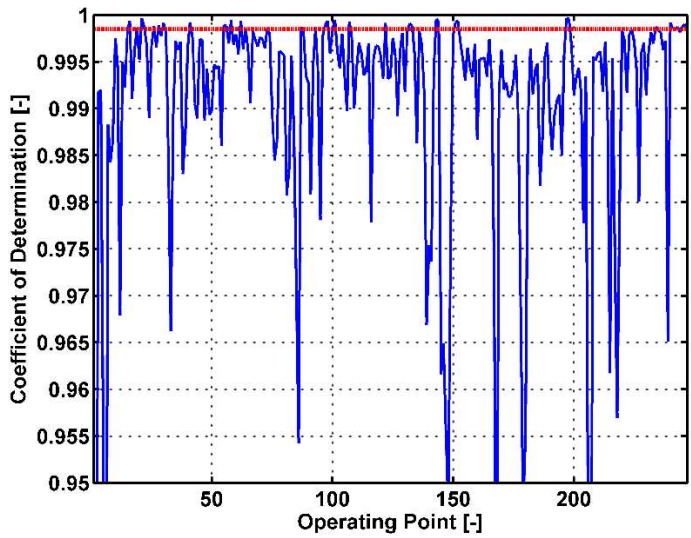

Figure 7. Coefficient of Determination of Lognormal Fitting. The red line represents the

95% confidence level (0.9985).

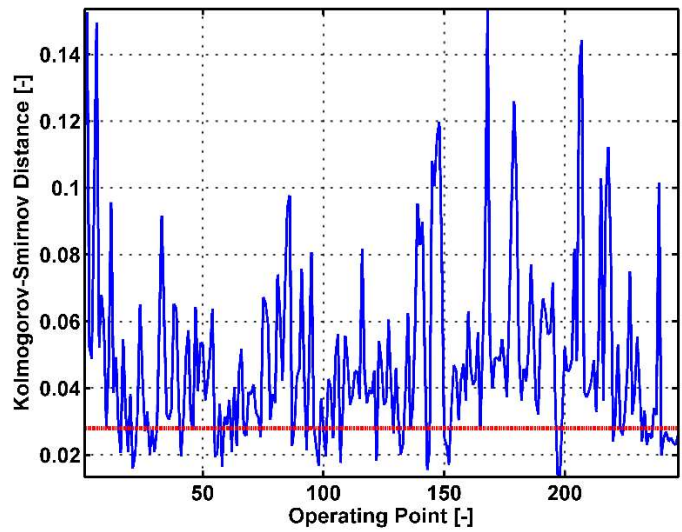

Figure 8. Kolmogorov-Smirnov Distance of Lognormal Fitting

*Mixed Lognormal Distribution*

In this section, similar steps to the previous section are undertaken to evaluate the accuracy of a mixed lognormal fitting of KI. The mixed lognormal distribution, unlike the lognormal distribution, can account for subgroups in the fitted dataset. For instance, across the 1116 cycles of a single operating point, some fraction of the cycle might be considered knocking and others non-knocking, each lognormally distributed with distinct means and standard deviation. Although the lognormal distribution fitting will fail to capture the distinct distributions of each subgroup, the mixed lognormal distribution will individually account for each distribution subsequently fitting the whole distribution more accurately. The PDF and CDF of a mixed lognormal distribution are given by Equations (8) and (9) respectively,

$$f(x) = a \cdot \frac{1}{x\sqrt{2\pi\sigma_1^2}} e^{-\frac{(\ln(x)-\mu_1)^2}{2\sigma_1^2}} + (1-a)$$

$$\cdot \frac{1}{x\sqrt{2\pi\sigma_2^2}} e^{-\frac{(\ln(x)-\mu_2)^2}{2\sigma_2^2}} \qquad (8)$$

$$F(x) = a \cdot \left(\frac{1}{2} + \frac{1}{2}\mathrm{erf}\left(\frac{(\ln(x)-\mu_1)^2}{2\sigma_1^2}\right)\right) + (1-a) \qquad (9)$$

$$\cdot \left(\frac{1}{2} + \frac{1}{2}\mathrm{erf}\left(\frac{(\ln(x)-\mu_2)^2}{2\sigma_2^2}\right)\right)$$

where $\mu_i$ and $\sigma_i$ are the mean and standard deviation of the variable's subgroup's natural logarithm, and $a$ represents the fractional size of the first subgroup.

The estimation of the parameters ($a, \mu_1, \mu_2, \sigma_1$, and $\sigma_2$) is accomplished using an Expectation-Maximization (EM) algorithm. Unlike the case for a lognormal distribution, the MLE for the variables in Equation (9) cannot be analytically resolved. The function *gmdistribution* on MATLAB is leveraged to determine the fitting variable.

The fitting accuracy of the mixed lognormal distribution is illustrated in Figure 9, which again shows three levels of fitting accuracy used to indicate the fitting accuracy of the worst cases, best cases, and median cases. Once again, the blue lines represent the PDF of the KI measurements, the green lines represent the empirical CDF, and the dotted red lines represent the resulting estimation of the PDF and CDF from the mixed lognormal fitting of KI (using $\hat{a}, \widehat{\mu_1}, \widehat{\mu_2}, \widehat{\sigma_1}$, and $\widehat{\sigma_2}$ from the EM estimation). The empirical distribution of KI is better captured by the mixed lognormal distribution. This is evident from both the $R^2$ and KS distance values evaluated in all three cases in Figure 9. Similar to the previous section, the fitting accuracy across the 247 operating points considered was seemingly random; Figure 9 is used to illustrate the range of fitting accuracy achieved with a mixed lognormal fit of the KI distributions.

The fitting metrics ($R^2$ and KS distance) are once again compared with the fitting accuracy observed from 10,000 randomly generated (mixed lognormally distributed) data points. The 5[th] percentile $R^2$ value was determined to be 0.9995 and the 95[th] percentile KS distance was found to be 0.017. Both represent higher fitting accuracy thresholds than the lognormal distribution, which is expected since an additional 3 parameters are introduced. The fitting accuracy of the 10,000 randomly generated points provides a baseline for the evaluation of the fitting accuracy of KI. If knock or knock intensity behaves as a mixed lognormally distributed random variable, the fitting accuracy (reflected by $R^2$ and KS distance) would yield similar results as that of the 10,000 mixed lognormally distributed data.

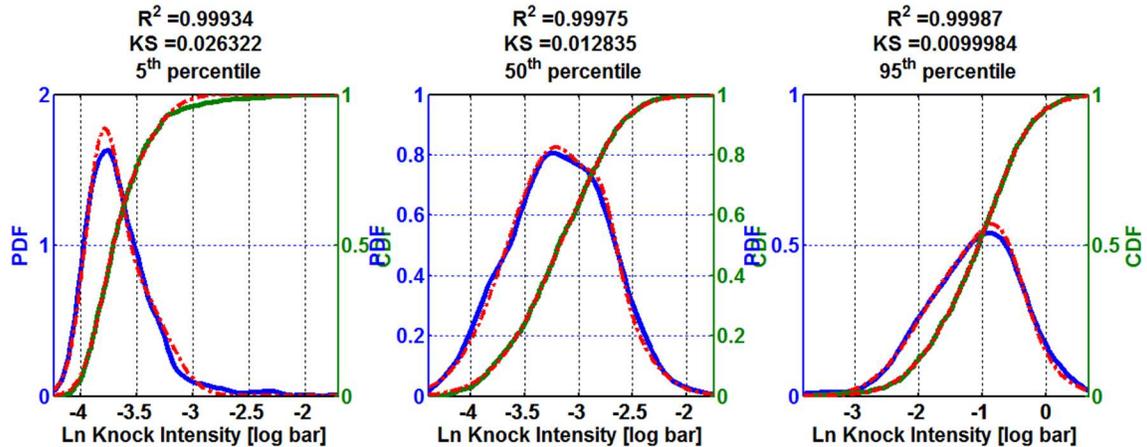

Figure 9. Mixed lognormal fitting of KI with varying degrees of fitting accuracy

As illustrated by Figure 10 and Figure 11, the fitting accuracy observed across the 247 operating points generally exceeds the fitting threshold. These results suggest that a mixed lognormal distribution behaves much better than the lognormal distribution and can be used to accurately capture the distribution of KI over a wide range of operating conditions for both single fuel and dual fuel engine configurations. This characterization of knock could be used to improve knock models in higher-fidelity engine models and also in more advanced knock control methods. One example of a stochastic control framework that could leverage this knock distribution characterization is given in the next section.

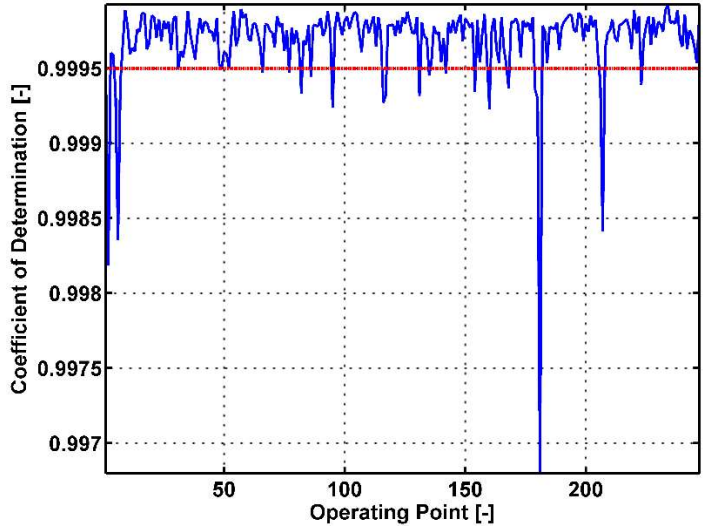

Figure 10. Coefficient of determination of mixed lognormal fitting.

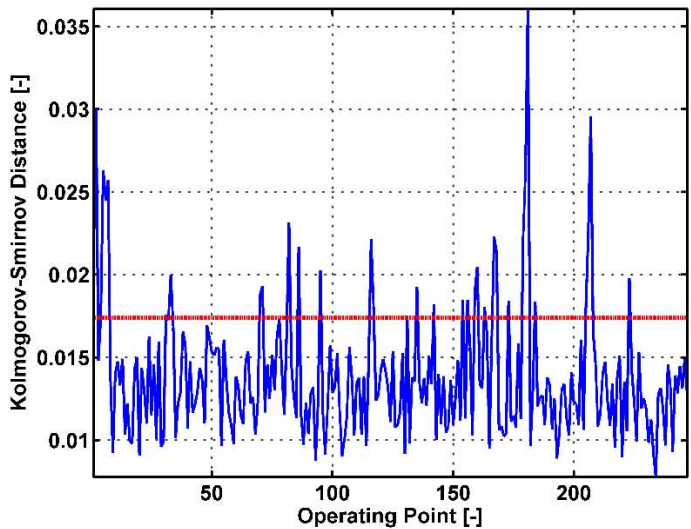

Figure 11. Kolmogorov-Smirnov Distance of mixed lognormal fitting.

**Stochastic Knock Control Framework**

Since KI can be captured as a mixed lognormal distribution, this can be leveraged in the control approach. Accurately determining the distribution of KI over a range of operating conditions could enable improved knock control strategies. For example, Figure 12 illustrates the evolution of KI distribution with varying spark timing. Each of these distributions corresponds to a certain level of knocking condition. The distribution in blue indicates a very low knocking condition, and the red most line indicates the distribution of a highly knocking condition. In each of the cases, the solid lines represent the experimental distribution, and the dotted red lines represent the mixed-lognormal fit of the distribution. Once again, the natural logarithm of KI plotted for better illustration (the distribution of KI is largely skewed to the left).

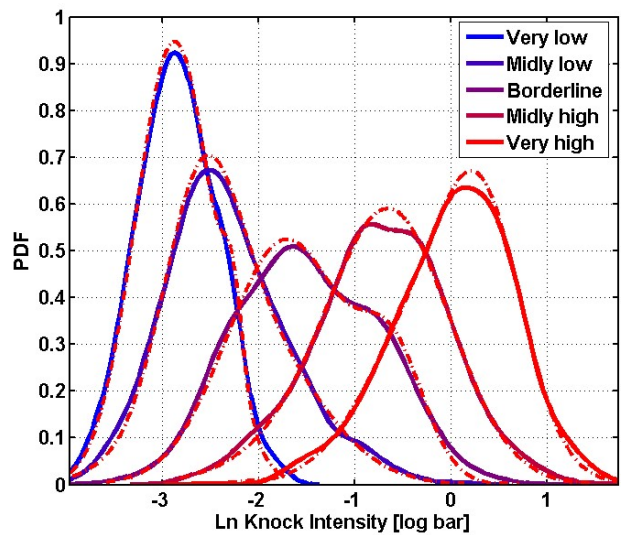

Figure 12. Evolution of KI distribution based on knock level

The proposed knock control approach would determine whether the measured KIs are indicative of a knocking condition or not. A-priori knowledge of how KIs are distributed with different levels of knock severity could be leveraged to develop the a-posteriori estimation of the knock state.

For example, the five states illustrated in Figure 12 could be considered to be exclusive states of knock. The knock state is either very low, mildly low, borderline, mildly high, or very high. The states can be defined as $M_{i\ for}$ i=1 to 5. The probability that the current case is either one of these states based on the measurements of KI ($x$) is represented as $P(M_j|x)$. Using Bayes' theorem, it follows

$$P(M_j|x) = \frac{P(x|M_j)P(M_j)}{P(x)} = \frac{P(x|M_j)P(M_j)}{\sum_{i=1}^{5} P(x|M_j)P(M_j)} \qquad (10)$$

where $P(x|M_j)$ can be simply calculated from a-priori distribution model corresponding to each state. The a-priori probability of each states $P(M_j)$ must be pre-determined. A plausible assumption is that without any measurement, each state is equally likely to be the current state ($P(M_j) = 1/(number\ of\ states)$). Alternatively, $P(M_j)$ can be adapted based on previous measurements.

Using the Bayesian approach, with each KI measurement, the likelihood of being in either knocking states can be analytically determined and used as an indicator for the spark timing or fuel-split control. When sets of KI measurements suggest a low knock state, the spark timing can be advanced accordingly and the fuel split (on a dual fuel engine) can be adjusted to preserve the high RON fuel. Alternatively, when the measurements suggest the high knock state is more likely to be the current state, spark timing can be retarded or more of a higher RON fuel can be injected.

A simplified control framework is illustrated in Figure 13, in which the spark advance is determined based on the estimated probability of each knock states. The measurements of KI are used to determine the probabilities that the current operating condition is one of the considered 5 knocking states. For example, $P(M_5|KI)$ is the probability that state 5 is the current state given the measurements $KI$. State $M_1$ through $M_5$ represent the "very low" to "very high" knocking states respectively (as illustrated in Figure 12).

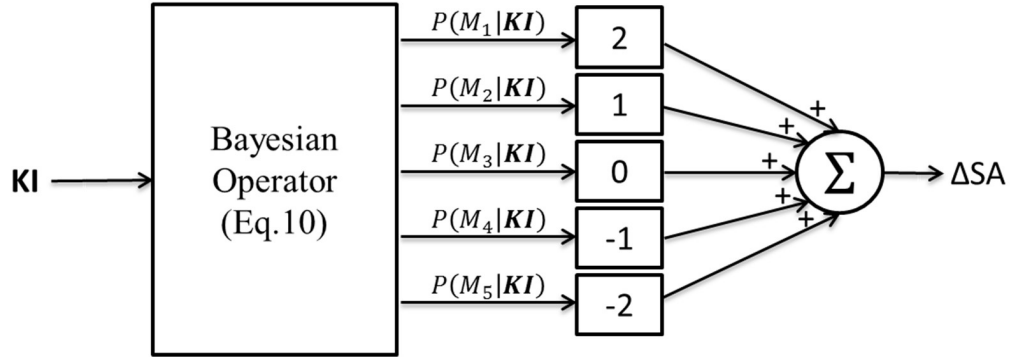

Figure 13. Illustration of stochastic knock controller framework

In the extreme cases, if the set of measurements suggest $P(M_5|KI) = 1$ (very high knocking states), then spark timing is retarded by 2°CA; if the set of measurements suggest $P(M_1|KI) = 1$ (very low knocking states), then spark timing is advanced by 2°CA; and, if the set of measurements suggest $P(M_3|KI) = 1$ (borderline knocking states), then the spark timing is kept constant. Otherwise, the spark advance will be a weighted sum based on the probability of each state.

Note,

$$\sum_{i=1}^{5} P(M_i|KI) = 1 \qquad (11)$$

where $0 \leq P(M_i|KI) \leq 1$ for $i = 1$ to 5.

The control framework can be adapted to feature as many (or few) knock states as desired. Furthermore, the spark advance associated with each state can also be varied. Lastly, the size of the set of measurements of KI used to calculate the probability of the states is also a parameter that can be adjusted based on the desired performance.

A control strategy that relies on the a-priori information of knock intensity distribution can leverage all measurements of KI to yield better performance for knock control on SI engines. Such a control approach allows the engine to operate close to the knock limit (thereby improving thermal efficiency), and provides an alternative to the conventionally used (conservative) knock control strategy. A stochastic control framework leveraging the mixed lognormal distribution of KI is proposed. Future work will seek to evaluate the performance of the control strategy and determine the potential gains in fuel efficiency that can be achieved by implementing a less conservative and more informed knock control approach.

**Conclusion**

Cycle-to-cycle control of knock is an essential aspect of the SI engine. A proper evaluation of KI is necessary for effective control of the ignition and fueling strategy to assure that knock is avoided while simultaneously maximizing fuel efficiency. KI is shown to be cyclically independent throughout the operating points tested; as such a deterministic knock control approach that adjusts the spark timing based on the previous cycle KI seems counterintuitive. Furthermore, engine manufacturers are forced to employ highly conservative knock control strategies to deal with the random nature of knock. Improvement to the knock control strategy can be achieved by leveraging the statistical properties of KI, which translates directly to better fuel efficiency.

In this study, it is shown that KI conforms to a mixed lognormal distribution over a wide range of operating points. The distribution of KI measurements at 247 operating points is fitted with a mixed lognormal CDF and shown to yield fitting accuracies greater than the 95% confidence threshold. The characterization of KI as a mixed lognormal event suggests that there are sets of cycles (ex. knocking vs. non-knocking) for the same operating point, which exhibit distinct properties and should be modeled separately.

Lastly, a stochastic knock control framework is proposed that leverages the mixed lognormal distribution of knock intensity to adjust the spark timing in order to avoid knock while minimizing fuel efficiency losses. The control strategy uses a-priori information (the distribution of knock intensity) to identify the likelihood of the knocking propensity; as such, a more informed control of the spark timing can be achieved. The number of knock states required to achieve a desired level of control may vary from different engines, but the proposed approach can be applied across engine architectures.

The knock intensity distribution characterization and the associated knock control framework are developed based on experimental data generated from a dual-fuel engine; nevertheless, the applicability of the distribution model and the control framework are not limited to this particular engine or such engine architecture. The findings of the paper highlight intrinsic properties associated with knock and illustrate a framework that can improve the knock control strategy on modern combustion engines.

**Acknowledgements**

The submitted manuscript has been created by UChicago Argonne, LLC, Operator of Argonne National Laboratory ("Argonne"). Argonne, a U.S. Department of Energy Office of Science laboratory, is operated under Contract No. DE-AC02-06CH11357. The U.S. Government retains for itself, and others acting on its behalf, a paid-up nonexclusive, irrevocable worldwide license in said article to reproduce, prepare derivative works, distribute copies to the public, and perform publicly and display publicly, by or on behalf of the Government. The Department of Energy will provide public access to these results of federally sponsored research in accordance with the DOE Public Access Plan. http://energy.gov/downloads/doe-public-access-plan

Argonne National Laboratory's work was supported by the U.S. Department of Energy, Office of Energy Efficiency and Renewable Energy, Office of Vehicle Technology under contract DE-AC02-06CH11357.

**References**

1. Haskell, W. and Bame, J., "Engine Knock -An End-Gas Explosion," SAE Technical Paper 650506, 1965, https://doi.org/10.4271/650506.

2. X. Zhen, Y. Wang, S. Xu, Y. Zhu, C. Tao, T. Xu and M. Song, "The engine knock analysis--An overview," *Applied Energy,* vol. 92, pp. 628-636, 2012.

3. L. Guzzella and C. Onder, "Control of Engine Systems: Engine Knock," in *Introduction to Modeling and Control of Internal Combustion Engine Systems*, Berlin, Springer Berlin, 2014, pp. 199-209.

4. Ayala, F., Gerty, M., and Heywood, J., "Effects of Combustion Phasing, Relative Air-fuel Ratio, Compression Ratio, and Load on SI Engine Efficiency," SAE Technical Paper 2006-01-0229, 2006, https://doi.org/10.4271/2006-01-0229.

5. Baranski, J., Anderson, E., Grinstead, K., Hoke, J. et al., "Control of Fuel Octane for Knock Mitigation on a Dual-Fuel Spark-Ignition Engine," SAE Technical Paper 2013-01-0320, 2013, https://doi.org/10.4271/2013-01-0320.

6. D. R. Cohn, L. Bromberg and J. B. Heywood, "Direct Injection Ethanol Boosted Gasoline Engines: Biofuel Leveraging For Cost Effective Reduction of Oil Dependence and CO2 Emissions. April 20, 2005," *Plasma Science and Fusion Center, Massachusetts Institute of Technology,* 2005.

7. R. Daniel, C. Wang, H. Xu, G. Tian and D. Richardson, "Dual-injection as a knock mitigation strategy using pure ethanol and methanol," *SAE International Journal of Fuels and Lubricants,* vol. 5, pp. 772-784, 2012.

8. Viollet, Y., Abdullah, M., Alhajhouje, A., and Chang, J., "Characterization of High Efficiency Octane-On-Demand Fuels Requirement in a Modern Spark Ignition Engine with Dual Injection System," SAE Technical Paper 2015-01-1265, 2015, https://doi.org/10.4271/2015-01-1265.

9. Naber, J., Blough, J., Frankowski, D., Goble, M. et al., "Analysis of Combustion Knock Metrics in Spark-Ignition Engines," SAE Technical Paper 2006-01-0400, 2006, https://doi.org/10.4271/2006-01-0400.

10. J. M. Spelina, P. a. J. C. Jones and J. Frey, "Characterization of knock intensity distributions: Part 1: statistical independence and scalar measures," *Proceedings of the Institution of Mechanical Engineers, Part D: Journal of Automobile Engineering,* vol. 228, pp. 117-128, 2014.

11. J. M. Spelina, P. a. J. C. Jones and J. Frey, "Characterization of knock intensity distributions: Part 2: parametric models," *Proceedings of the Institution of Mechanical Engineers, Part D: Journal of Automobile Engineering,* vol. 227, pp. 1650-1660, 2013.

12. A. A. Stotsky, "Statistical engine knock modelling and adaptive control," *Proceedings of the Institution of Mechanical Engineers, Part D: Journal of Automobile Engineering,* vol. 222, pp. 429-439, 2008.

13. J. C. P. Jones, J. Frey, K. R. Muske and D. J. Scholl, "A cumulative-summation-based stochastic knock controller," *Proceedings of the Institution of Mechanical Engineers, Part D: Journal of Automobile Engineering,* vol. 224, pp. 969-983, 2010.

14. J. C. P. Jones, J. Frey and K. R. Muske, "A fast-acting stochastic approach to knock control," *IFAC Proceedings Volumes,* vol. 42, pp. 16-23, 2009.

15. J. C. P. Jones, J. M. Spelina and J. Frey, "Likelihood-based control of engine knock," *IEEE Transactions on Control Systems Technology,* vol. 21, pp. 2169-2180, 2013.

16. Penese, M., Damasceno, C., Bucci, A., and Montanari, G., "Sigma® on knock phenomenon control of Flexfuel engines," SAE Technical Paper 2005-01-3990, 2005, https://doi.org/10.4271/2005-01-3990.

17. N. Cavina, G. Po and L. Poggio, "Ion current based spark advance management for maximum torque production and knock control," in *Proc. Eighth Biennial ASME Conference on Engineering Systems Design and Analysis*, 2006.

18. G. Zhu, I. Haskara and J. Winkelman, "Stochastic limit control and its application to spark limit control using ionization feedback," in *American Control Conference, 2005. Proceedings of the 2005*, 2005.

19. Pamminger, M., Sevik, J., Scarcelli, R., Wallner, T. et al., "Influence of Compression Ratio on High Load Performance and Knock Behavior for Gasoline Port-Fuel Injection, Natural Gas Direct Injection and Blended Operation in a Spark Ignition Engine," SAE Technical Paper 2017-01-0661, 2017, doi:10.4271/2017-01-0661.

20. Pamminger, M., Sevik, J., Scarcelli, R., Wallner, T. et al., "Evaluation of Knock Behavior for Natural Gas - Gasoline Blends in a Light Duty Spark Ignited Engine," SAE Int. J. Engines 9(4):2016, doi:10.4271/2016-01-2293.

21. Hall, C., Sevik, J., Pamminger, M., and Wallner, T., "Hydrocarbon Speciation in Blended Gasoline-Natural Gas Operation on a Spark-Ignition Engine," SAE Technical Paper 2016-01-2169, 2016, https://doi.org/10.4271/2016-01-2169.

22. Baral, B. and Raine, R., "Knock in a Spark Ignition Engine Fuelled with Gasoline-Kerosene Blends," SAE Technical Paper 2008-01-2417, 2008, doi:10.4271/2008-01-2417

23. Mittal, V., Revier, B., and Heywood, J., "Phenomena that Determine Knock Onset in Spark-Ignition Engines," SAE Technical Paper 2007-01-0007, 2007, doi:10.4271/2007-01-0007

24. Kassa, M. *Analysis and control of compression-ignition and spark-ignited engines operating with dual-fuel combustion strategy*. Dissertation, Illinois Institute of Technology. Ann Arbor: ProQuest/UMI, 2017.

## List of Table Captions

Table 1. Single Cylinder Engine Specification

Table 2. Summary of Operating Parameters

## List of Figure Captions

Figure 1. Illustration of the pressure trace and filtered pressure trace (3-25kHz) for a sample combustion cycle. The red line indicates spark timing; the blue lines represent the knock window.

Figure 2. Filtered pressure trace amplitude evolution with spark timing
(E10 - 1500RPM / 7bar IMEP)

Figure 3. Evolution of average knock intensity with spark timing. The error bars represent the 5th and 95th percentile. (E10 - 1500RPM / 7bar IMEP)

Figure 4. Normalized autocorrelation function of KI values

Figure 5. PDF and CDF representation of sample KI values

Figure 6. Lognormal fitting of KI with varying degrees of fitting accuracy

Figure 7. Coefficient of Determination of Lognormal Fitting. The red line represents the 95% confidence level (0.9985).

Figure 8. Kolmogorov-Smirnov Distance of lognormal fitting

Figure 9. Mixed lognormal fitting of KI with varying degrees of fitting accuracy

Figure 10. Coefficient of determination of mixed lognormal fitting

Figure 11. Kolmogorov-Smirnov Distance of mixed lognormal fitting

Figure 12. Evolution of KI distribution based on knock level

Figure 13. Illustration of stochastic knock controller framework